\documentclass[10pt,conference,letterpaper]{IEEEtran}
\usepackage{graphicx}
\usepackage[utf8]{inputenc}
\usepackage{amsmath}

\usepackage{multirow}
\usepackage{amsfonts}
\usepackage{booktabs}
\usepackage{hyperref}
\usepackage[noadjust]{cite}
 \usepackage[svgnames,dvipsnames]{xcolor}
\newcommand{\osqrt}[1]{\mathop{\sqrt{#1}\!\!\!\!\!\!\circ\,\,\,}} 
\newcommand\myhead[2]{{\parbox{#1}{\centering \textbf{#2}}}}
\usepackage{listings}
\lstdefinelanguage{scala}{
  frame=tb,
  alsoletter={@=>},
  morekeywords={
  abstract, case, catch, class, def, do, else, extends, false, final, finally,
  for, if, implicit, import, match, new, null, object, override, package,
  private, protected, requires, return, sealed, super, this, throw, trait, try,
  true, type, val, var, while, with, yield, domain, postcondition,
  precondition,invariant, constraint, assert, forAll,  _, return, @generator,
  ensure, require, ensuring,=>, Real, certainly, possibly, certify, errorBound,
  assertBound
  },
  sensitive=true,
  morecomment=[l]{//},
  morecomment=[s]{/*}{*/},
  commentstyle=\color{gray},
  showstringspaces=false,
  columns=fullflexible,
  mathescape=true,
  numberstyle=\tiny,
  basicstyle=\footnotesize\ttfamily,
  numbersep=5pt,
  stepnumber=2,
  numbers=none,                   
  morestring=[b]"
}
\usepackage{color}
\usepackage{threeparttable}
\usepackage[T1]{fontenc}  
\usepackage[scaled=0.80]{beramono}  

\definecolor{black}{rgb}{0,0,0}

\newcommand{\tlf}{\tilde{f}}
\newcommand{\tlx}{\tilde{x}}

\begin{document}
	\title{On Sound Relative Error Bounds for \\Floating-Point Arithmetic}

	\author{\IEEEauthorblockN{Anastasiia Izycheva and Eva Darulova}
	\IEEEauthorblockA{Max Planck Institute for Software Systems,
	Saarland Informatics Campus, Germany\\
	Email: izycheva@mpi-sws.org, eva@mpi-sws.org}}

	\maketitle

	\begin{abstract}
		State-of-the-art static analysis tools for verifying finite-precision
		code compute worst-case absolute error bounds on numerical errors. These are,
		however, often not a good estimate of accuracy as they do not take into
		account the magnitude of the computed values.
		Relative errors, which compute errors relative to the value's magnitude, are
		thus preferable. While today's tools do report relative error bounds, these
		are merely computed via absolute errors and thus not necessarily tight or
		more informative. Furthermore, whenever the computed value is close to zero
		on part of the domain, the tools do not report any relative error estimate
		at all. Surprisingly, the quality of relative error bounds computed by
		today's tools has not been systematically studied or reported to date.

		In this paper, we investigate how state-of-the-art static techniques for computing
		sound absolute error bounds can be used, extended and combined for the
		computation of relative errors.
		Our experiments on a standard benchmark set show that computing relative errors
		\emph{directly}, as opposed to via absolute errors, is often beneficial and
		can provide error estimates up to six orders of magnitude tighter, i.e. more
    accurate.
    We also show that interval subdivision, another commonly used
    technique to reduce over-approximations, has less benefit when computing relative
    errors directly, but it can help to alleviate the effects of the inherent issue
    of relative error estimates close to zero.

	\end{abstract}

\section{Introduction}\label{sec:intro}


Numerical software, common in embedded systems or scientific computing, is
usually designed in real-valued arithmetic, but has to be implemented in
finite precision on digital computers. Finite precision, however, introduces
unavoidable roundoff errors which are individually usually small, but which can
accumulate and affect the validity of computed results.
It is thus important to compute \emph{sound} worst-case roundoff error bounds
to ensure that results are accurate enough - especially for
safety-critical applications. Due to the unintuitive nature of finite-precision
arithmetic and its discrepancy with continuous real arithmetic,
\emph{automated} tool support is essential.

One way to measure worst-case roundoff is
\emph{absolute error}:
\begin{equation}
err_{\text{abs}}=\max_{x\in I}\left|f(x)-\hat{f}(\hat{x})\right|
\label{eqn:abserr}
\end{equation}
where $f$ and $x$ denote a possibly multivariate real-valued
function and variable respectively, and $\hat{f}$ and $\hat{x}$ their
finite-precision counter-parts.
Note that absolute roundoff errors are only meaningful on a restricted domain,
as for unrestricted $x$ the error would be unbounded in general. In this paper,
we consider interval constraints on input variables, that is for each input variable
$x \in I = [a, b]$, $a, b \in \mathbb{R}$.

Furthermore, we focus on floating-point arithmetic, which is a common choice for
many finite-precision programs, and for which several tools now exist
that compute absolute errors fully automatically for nonlinear straight-line
code~\cite{fluctuat,fptaylor,rosa,Magron2015}.

Absolute errors are, however, not always an adequate measure of result quality.
Consider for instance an absolute error of $0.1$. Whether this error is small
and thus acceptable for a computation depends on the application as well
as the magnitude of the result's value: if $|f(x)| \gg0.1$, then the
error may be acceptable, while if $|f(x)| \approx0.1$ we should probably revise
the computation or increase its precision.
\emph{Relative error} captures this relationship:
\begin{equation}
err_{\text{rel}}=\max_{x\in I}\left|\frac{f(x)-\hat{f}(\hat{x})}{f(x)}\right|
\label{eqn:relerr}
\end{equation}
Note that the input domain needs to be restricted also for relative errors.

Today's static analysis tools usually report absolute as well as relative
errors. The latter is, however, computed via absolute errors. That is, the tools
first compute the absolute error and then divide it by the largest function
value:
\begin{equation}
err_{\text{rel\_approx}}=\frac{\max_{x\in I}\left|f(x)-\hat{f}(\hat{x})\right|}{\min_{x\in I}\left|f(x)\right|}
\label{eqn:relerr_via_abs}
\end{equation}
Clearly,~\autoref{eqn:relerr} and~\autoref{eqn:relerr_via_abs} both compute
sound relative error bounds, but $err_{\text{rel\_approx}}$ is an
over-approximation due to the loss of correlation between the nominator and
denominator. Whether this loss of correlation leads to coarse error bounds in
practice has, perhaps surprisingly, not been studied yet in the context of
automated sound error estimation.

Beyond curiosity, we are interested in the automated computation of relative
errors for several reasons. First, relative errors are more informative and
often also more natural for user specifications. Secondly, when computing sound
error bounds, we necessarily compute over-approximations. For absolute errors,
the over-approximations become bigger for larger input ranges, i.e. the error
bounds are less tight. Since relative errors consider the range of the
expression, we expect these over-approximations to be smaller, thus making
relative errors more suitable for modular verification.

One may think that computing relative errors is no more challenging than
computing absolute errors; this is not the case for two reasons. First, the
complexity of computing relative errors is higher (compare~\autoref{eqn:abserr}
and~\autoref{eqn:relerr}) and due to the division, the expression is now
nonlinear even for linear $f$. Both complexity and nonlinearity have already
been challenging for absolute errors computed by automated tools, usually
leading to coarser error bounds.
Furthermore, whenever the range of $f$ includes zero, we face an inherent
division by zero. Indeed, today's static analysis tools report no relative
error for most standard benchmarks for this reason.

Today's static analysis tools employ a variety of different strategies (some
orthogonal) for dealing with the over-approximation of worst-case absolute
roundoff errors due to nonlinear arithmetic:
the tool Rosa uses a forward dataflow analysis with a linear abstract domain
combined with a nonlinear decision procedure~\cite{rosa}, Fluctuat augments a
similar linear analysis with interval subdivision~\cite{fluctuat}, and FPTaylor
chooses an optimization-based approach~\cite{fptaylor} backed by a branch-and-bound
algorithm.

In this paper, we investigate how today's methods can be used, extended and
combined for the computation of relative errors. To the best of our knowledge,
this is the first systematic study of fully automated techniques for the
computation of relative errors.
We mainly focus on the issue of computing tight relative error bounds
and for this extend the optimization based approach
for computing absolute errors to computing relative errors \emph{directly} and
show experimentally that it often results in tighter error bounds, sometimes by
up to six orders of magnitude.
We furthermore combine it with interval subdivision
(we are not aware of interval subdivision being applied to this approach before),
however, we note that, perhaps surprisingly, the benefits are rather modest.

We compare this direct error computation to forward analysis which computes relative
errors via absolute errors on a standard benchmark set, and note
that the latter outperforms direct relative error computation only on a single univariate
benchmark. On the other hand, this approach clearly benefits from interval
subdivision.

We also observe that interval subdivision is beneficial for dealing with the
inherent division by zero issue in relative error computations. We propose a
practical (and preliminary) solution, which reduces the impact of potential
division-by-zero's to small subdomains, allowing our tool to compute relative
errors for the remainder of the domain. We demonstrate on our benchmarks that this approach
allows our tool to provide more useful information than state-of-the-art tools.

\subsubsection*{Contributions}
\begin{itemize}
  \item We extend an optimization-based approach~\cite{fptaylor} for bounding
  absolute errors to relative errors and thus provide the first feasible and
  fully automated approach for computing relative errors \emph{directly}.

  \item We perform the first experimental comparison of different techniques for
  automated computation of sound relative error bounds.

  \item We show that interval subdivision is beneficial for reducing the
  over-approximation in absolute error computations, but less so for relative
  errors computed directly.

  \item We demonstrate that interval subdivision provides a practical solution
  to the division by zero challenge of relative error computations for certain benchmarks.


\end{itemize}
We have implemented all techniques within the tool Daisy~\cite{daisysrc},
which is available at \url{https://github.com/malyzajko/daisy}.

\section{Background}\label{sec:background}
We first give a brief overview over floating-point arithmetic as well as
state-of-the-art techniques for automated sound worst-case absolute roundoff
error estimation.

\subsection{Floating-Point Arithmetic}\label{sec:rosa}

  The error definitions in~\autoref{sec:intro} include a finite-precision
  function $\hat{f}(\hat{x})$ which is highly irregular and discontinuous and
  thus unsuitable for automated analysis. We abstract it following the
  floating-point IEEE~754 standard~\cite{ieee754}, by replacing every
  floating-point variable, constant and operation by:
  \begin{equation}
  \begin{gathered}
  x \odot y = (x \circ y)(1+e) + d,\\
  \tilde x=x(1+e)+d \quad\quad \quad\osqrt{x} = \sqrt{x}(1+e)  + d
  \end{gathered}
  \label{eq:fp_model}
  \end{equation}
  where $\odot \in \{\oplus,\ominus,\otimes,\oslash\}$ and $\circ \in
  \{+,-,\times,/\}$ are floating-point and real arithmetic operations,
  respectively. $e$ is the relative error introduced by rounding at each
  operation and is bounded by the so-called machine epsilon $|e| \le
  \epsilon_M$.
  Denormals (or subnormals) are values with a special representation
  which provide gradual underflow. For these, the
  roundoff error is expressed as an absolute error $d$ and is bounded by $\delta_M$,
  (for addition and subtraction $d=0$).
  This abstraction is valid in the absence of overflow and invalid operations
  resulting in Not a Number (NaN) values. These values are detected separately
  and reported as errors.
  In this paper, we consider double precision floating-point arithmetic
  with $\epsilon_M = 2^{-53}$ and $\delta_M=2^{-1075}$.
  Our approach is parametric in the precision, and thus applicable to
  other floating-point types as well.

  Using this abstraction we replace $\hat{f}(\hat{x})$ with a function $\tilde
  f(x,e,d)$, where $x$ are the input variables and $e$ and $d$ the roundoff
  errors introduced for each floating-point operation. In general, $x, e$ and
  $d$ are vector-valued, but for ease of reading we will write them without
  vector notation. Note that our floating-point abstraction is real-valued. With
  this abstraction, we and all state-of-the-art analysis tools approximate
  absolute errors as:
  \begin{equation}
   err_{abs} \leq \max_{x\in I,|e_i|\leq\epsilon_M,|d_i|\leq\delta}\left|f(x) - \tilde f(x,e,d)\right|
   \label{eq:abserr:appr}
  \end{equation}

\subsection{State-of-the-art in Absolute Error Estimation}
When reviewing existing automated tools for absolute roundoff error estimation,
we focus on their techniques for reducing over-approximations due to nonlinear
arithmetic in order to compute tight error bounds.

\emph{Rosa}~\cite{rosa} computes absolute error bounds using a forward data-flow
  analysis and a combination of abstract domains.
  Note that the magnitude of the absolute roundoff error at an arithmetic
  operation depends on the magnitude of the operation's value (this can easily
  be seen from~\autoref{eq:fp_model}), and these are in turn determined by the
  input parameter ranges. Thus, Rosa tracks two values for each intermediate
  abstract syntax tree node: a sound approximation of the range and the
  worst-case absolute error. The transfer function for errors uses the
  ranges to propagate errors from subexpressions and to compute the new roundoff
  error committed by the arithmetic operation in question.

  One may think that evaluating an expression in interval
  arithmetic~\cite{Moore1966} and interpreting the width of the resulting
  interval as the error bound would be sufficient. While this is sound,
  it computes too pessimistic error bounds, especially if we consider relatively
  large ranges on inputs: we cannot
  distinguish which part of the interval width is due to the input
  interval or due to accumulated roundoff errors. Hence, we need to compute
  ranges and errors separately.

  Rosa implements different range arithmetics with different accuracy-efficiency
  tradeoffs for bounding ranges and errors.
  To compute ranges, Rosa offers a choice between interval arithmetic, affine
  arithmetic~\cite{Figueiredo2004} (which tracks linear correlations between
  variables) and a combination of interval arithmetic with a nonlinear arithmetic
  decision procedure. The latter procedure first computes the range of an expression in
  standard interval arithmetic and then refines, i.e. tightens, it using
  calls to the nlsat~\cite{Jovanovic2012} decision procedure within the Z3 SMT
  solver~\cite{De-Moura2008}. For tracking errors, Rosa uses affine arithmetic;
  since roundoff errors are in general small, tracking linear correlations is in
  general sufficient.

\emph{Fluctuat}~\cite{fluctuat} is an abstract interpreter which separates errors similarly to Rosa
  and which uses affine arithmetic for computing both the ranges of variables
  and for the error bounds.
  In order to reduce the over-approximations introduced by affine arithmetic for
  nonlinear operations, Fluctuat uses interval subdivision. The user can
  designate up to two variables in the program whose input ranges will be
  subdivided into intervals of equal width. The analysis is performed separately
  and the overall error is then the maximum error over all subintervals.
  Interval subdivision increases the runtime of the analysis significantly,
  especially for multivariate functions, and the choice of which variables to
  subdivide and by how much is usually not straight-forward.

  \emph{FPTaylor}, unlike Daisy and Fluctuat, formulates
  the roundoff error bounds computation as an optimization problem, where the
  absolute error expression from~\autoref{eqn:abserr} is to be maximized,
  subject to interval constraints on its parameters. Due to the discrete nature
  of floating-point arithmetic, FPTaylor optimizes the continuous,
  real-valued abstraction (\autoref{eq:abserr:appr}). However, this expression is
  still too complex and features too many variables for optimization procedures
  in practice, resulting in bad performance as well as
  bounds which are too coarse to be useful (see~\autoref{sec:eval_nozero} for
  our own experiments).
  FPTaylor introduces the Symbolic Taylor approach, where the objective
  function of~\autoref{eq:abserr:appr} is simplified using a first order Taylor
  approximation with respect to $e$ and $d$:
   \begin{equation}
   \tilde{f}(x, e, d) = \tilde f(x,0,0) + \sum_{i =1}^{k}\frac{\partial \tilde f}{\partial e_i}(x, 0, 0)e_i + R(x,e, d),\\
    \label{eq:fptay:tildeexp}
    \end{equation}
    \begin{equation*}
   R(x,e,d) = \frac{1}{2}\sum_{i,j =1}^{2k}\frac{\partial^2 \tilde f}{\partial y_i\partial y_j}(x, p)y_iy_j + \sum_{i=1}^{k}\frac{\partial \tilde f}{\partial d_i}(x,0,0)d_i
   \label{eq:fptay:remainder}
   \end{equation*}
  where $y_1=e_1,\ldots,y_k=e_k,y_{k+1}=d_1,\ldots,y_{2k}=d_k$ and
  $p\in\mathbb{R}^{2k}$ such that $|p_i|\leq\epsilon_M$ for $i=1\ldots k$ and
  $|p_i|\leq\delta$ for $i=k+1\ldots 2k$.
  The remainder term $R$ bounds all higher order terms and ensures soundness
  of the computed error bounds.

  The approach is symbolic in the sense that the Taylor approximation is
  taken wrt. $e$ and $d$ only and $x$ is a symbolic argument.
  Thus, $f(x,0,0)$ represents the function point where all inputs $x$ remain
  symbolic and no roundoff errors are present, i.e. $e = d = 0$ and $f(x,0,0)=f(x)$.
  Choosing $e = d = 0$ as the point at which to perform the Taylor approximation
  and replacing $e_i$ with its upper bound $\epsilon_M$
  reduces the initial optimization problem to:
  \begin{equation}
  err_{abs}\leq \epsilon_M \max_{x\in I}\sum_{i=1}^{k}\left|\frac{\partial \tilde f}{\partial e_i}(x, 0,0)\right|+M_R
  \label{eq:fptay:fin}
  \end{equation}
  where $M_R$ is an upper bound for the error term $R(x,e,d)$
  (more details can be found in~\cite{fptaylor}). FPTaylor uses
  interval arithmetic to estimate the value of $M_R$ as the term is second
  order and thus small in general.

  To solve the optimization problem in~\autoref{eq:fptay:fin}, FPTaylor uses
  rigorous branch-and-bound optimization. Branch-and-bound is also used to
  compute the resulting real function $f(x)$ range, which is needed for instance
  to compute relative errors.
  Real2Float~\cite{Magron2015}, another tool, takes the same optimization-based approach,
  but uses semi-definite programming for the optimization itself.

\section{Bounding Relative Errors}\label{sec:rel-errors}

The main goal of this paper is to investigate how today's sound approaches for
computing absolute errors fare for bounding relative errors and whether it is
possible and advantageous to compute relative errors directly (and not via absolute
errors). In this section, we first concentrate on obtaining tight bounds in the presence
of nonlinear arithmetic, and postpone a
discussion of the orthogonal issue of division by zero to the next section.
Thus, we assume for now that the range of the function for which we want to
bound relative errors does not include zero, i.e. $0 \notin f(x)$ and $0 \notin
\tlf(\tlx)$, for $x, \tlx$ within some given input domain.
We furthermore consider $f$ to be a straight-line arithmetic expression.
Conditionals and loops have been shown
to be challenging~\cite{Darulova2017} even for absolute errors and we thus leave
their proper treatment for future work. We consider
function calls to be an orthogonal issue; they can be handled by inlining, thus
reducing to straight-line code, or require suitable summaries in postconditions,
which is also one of the motivations for this work.

The forward dataflow analysis approach of Rosa and Fluctuat does not easily
generalize to relative errors, as it requires intertwining the range and error
computation. Instead, we extend FPTaylor's approach to
computing relative errors directly (\autoref{sec:tay_approach}). We furthermore
implement interval subdivision (\autoref{sec:int_subdiv}), which is an
orthogonal measure to reduce over-approximation and experimentally evaluate
different combinations of techniques on a set of standard benchmarks
(\autoref{sec:eval_nozero}).

\subsection{Bounding Relative Errors Directly}\label{sec:tay_approach}
  The first strategy we explore is to compute relative errors directly,
  without taking the intermediate step through absolute errors. That is, we
  extend FPTaylor's optimization based approach and maximize the relative error
  expression using the floating-point abstraction from~\autoref{eq:fp_model}:
  \begin{equation}
    \max |\tilde g(x,e,d)|=
    \max_{x\in I,|e_i|\leq\epsilon_M,|d_i|\leq\delta} \left|\frac{f(x)-\tlf(x,e,d)}{f(x)}\right|
    \label{eqn:g}
  \end{equation}
  The hope is to preserve more correlations between variables in the nominator
  and denominator and thus obtain tighter and more informative relative error
  bounds.

  We call the optimization of~\autoref{eqn:g} without simplifications the \emph{naive
  approach}. While it has been mentioned previously that this approach does not
  scale well~\cite{fptaylor}, we include it in our experiments (in~\autoref{sec:eval_nozero}) nonetheless, as
  we are not aware of any concrete results actually being reported.
  As expected, the naive approach returns error bounds which are so large that
  they are essentially useless.

  We thus simplify $\tilde g(x, e, d)$ by applying the Symbolic
  Taylor approach introduced by FPTaylor~\cite{fptaylor}.
  As before, we take the Taylor approximation around the point $(x,0, 0)$, so that
  the first term of the approximation is zero as before:
  $\tilde g(x,0, 0) = 0$.
  We obtain the following optimization problem:
  $$
  \max_{x\in I,|e_i|\leq\epsilon_M,|d_i|\leq\delta}\sum_{i =1}^{k}\left|\frac{\partial \tilde g}{\partial e_i}(x,0, 0)e_i\right| + M_R
  $$
  where $M_R$ is an upper bound for the remainder term $R(x,e,d)$.
  Unlike in~\autoref{eq:fptay:fin} we do not pull the
  factor $e_i$ for each term out of the absolute value, as we plan to compute tight bounds for
  mixed-precision in the future, where the upper bounds on all $e_i$ are not
  all the same (this change does not affect the accuracy for uniform precision computations).

  \subsubsection*{Computing Upper Bounds}
    The second order remainder $R$ is still expected to be small, so that we use interval
    arithmetic to compute a sound bound on $M_R$ (in our experiments $R$ is indeed small for
    all benchmarks except `doppler'). To bound the first order terms $\frac{\partial \tilde
    g}{\partial e_i}$ we need a sound optimization procedure to maintain overall
    soundness, which limits the available choices significantly.

    FPTaylor uses the global optimization tool Gelpia~\cite{gelpia}, which
    internally uses a branch-and-bound based algorithm. Unfortunately, we found
    it difficult to integrate because of its custom interface. Furthermore, we
    observed unpredictable behavior in our experiments (e.g. nondeterministic
    crashes and substantially varying running times for repeated runs on
    the same expression).

    Instead, we use Rosa's approach which combines interval arithmetic with a
    solver-based refinement. Our approach is parametric in the solver and we
    experiment with Z3~\cite{De-Moura2008} and dReal~\cite{Gao2013}. Both
    support the SMT-lib interface, provide rigorous results, but are based on
    fundamentally different techniques. Z3 implements a complete decision
    procedure for nonlinear arithmetic~\cite{Jovanovic2012}, whereas dReal implements the framework of
    $\delta$-complete decision procedures. Internally, it is based on a
    branch-and-bound algorithm and is thus in principle similar to Gelpia's
    optimization-based approach.

    Note that the queries we send to both solvers are (un)satisfiability
    queries, and not optimization queries. For the nonlinear decision procedure
    this is necessary as it is not suitable for direct optimization, but
    the branch-and-bound algorithm in dReal is performing optimization
    internally. The reason for our roundabout approach for dReal is that while
    the tool has an optimization interface, it uses a custom input format and is
    difficult to integrate. We expect this approach to affect mostly
    performance, however, and not accuracy.


\subsection{Interval Subdivision}\label{sec:int_subdiv}

  The over-approximation committed by static analysis techniques
  grows in general with the width of the input intervals, and thus with the width of all
  intermediate ranges. Intuitively, the worst-case error which we consider is
  usually achieved only for a small part of the domain, over-approximating the
  error for the remaining inputs. Additionally the domain where worst-case
  errors are obtained may be different at each arithmetic operation.
  Thus, by subdividing the input domain we can usually obtain tighter error
  bounds. Note that interval subdivision can be applied to any error estimation
  approach. Fluctuat has applied interval subdivision to absolute error
  estimation, but we are not aware of a combination with the optimization-based
  approach, nor about a study of its effectiveness for relative errors.

  We apply subdivision to input variables and thus compute:
  \begin{equation}
	  \max_{k\in[1\dots m]}\left(\max_{x_j\in I_{jk}}\left|\tilde{g}(x,e,d)\right|\right)
  \end{equation}
  where $m$ is an number of subdivisions for each input interval.
  That is, for multivariate functions, we subdivide the input interval for each
  variable and maximize the error over the Cartesian product. Clearly, the
  analysis running time is exponential in the number of variables.
  While Fluctuat limits subdivisions to two user-designated variables
  and a user-defined number of subdivisions each, we choose to limit the total
  number of analysis runs by a user-specified parameter $p$.
  That is, given $p$, $m$ (the desired number of subdivisions for each
  variable), and $n$ (number of input variables), the first $\lfloor \log_{m}(p-n)\rfloor$
  variables' domains are subdivided $m$ times, with larger input domains subdivided first. The
  remaining variable ranges remain undivided.


\subsection{Implementation}

 We implement all the described techniques in the tool
 Daisy~\cite{daisysrc}. Daisy, a successor of Rosa~\cite{rosa}, is a source-to-source
 compiler which generates floating-point implementations from real-valued
 specifications given in the following format:

\begin{lstlisting}[label={lst:daisy:output}]
        def bspline3(u: Real): Real = {
          require(0 <= u && u <= 1)
             - u * u * u / 6.0
        }
\end{lstlisting}


  Daisy is parametric in the \emph{approach} (naive, forward dataflow analysis
  or optimization-based), the \emph{solver} used (Z3 or dReal) and the number
  of \emph{subdivisions} (including none). Any combination of these three
  orthogonal choices can be run by simply changing Daisy's input parameters.



  Furthermore, Daisy simplifies the derivative expressions in the optimization-based
  approach ($x + 0 =x, x\times1 =x$, etc.). Unsimplified expressions may affect the running time of the
  solvers (and thus also the accuracy of the error bounds), as we observed that
  the solvers do not necessarily perform these otherwise straight-forward
  simplifications themselves.

  Finally, to maintain soundness, we need to make sure that we do not introduce
  internal roundoff errors during the computation of error bounds. For this
  purpose we implement all internal computations in Daisy using infinite-precision
  rationals.

\section{Handling Division by Zero}\label{sec:div-by-zero}
%
An important challenge arising while computing relative errors is how to handle
potential divisions by zero. State-of-the-art tools today simply do not report
any error at all whenever the function range contains zero. Unfortunately, this
is not a rare occurrence; on a standard benchmark set for floating-point verification,
many functions exhibit division by zero (see~\autoref{tab:eval:div-by-zero} for
our experiments).

Note that these divisions by zero are \emph{inherent} to the expression which we
consider and are usually not due to over-approximations in the analysis. Thus, we
can only \emph{reduce} the effect of division by zero, but we cannot
eliminate it entirely. Here, we aim to reduce the domain for which we
cannot compute relative errors. Similar to how relative and absolute errors are
combined in the IEEE~754 floating-point model (\autoref{eq:fp_model}), we want
to identify parts of the input domain on which relative error computation is not possible
due to division by zero and compute absolute errors.
For the remainder of the input domain, we compute relative errors as before.




We use interval subdivision~(\autoref{sec:int_subdiv}), attempting
to compute relative errors (with one of the methods described before) on every
subdomain. Where the computation fails due to (potential) division by zero,
we compute the absolute error and report both to the user:

\begin{lstlisting}[label={lst:subdiv:output},basicstyle={\footnotesize\ttfamily}]
relError: 6.6614143807162e-16
On several sub-intervals relative error cannot be computed.
Computing absolute error on these sub-intervals.
For intervals (u -> [0.875,1.0]), absError: 9.66746937132909e-19
\end{lstlisting}
While the reported relative error bound is only sound for parts of the domain, we
believe that this information is nonetheless more informative than either no result at
all or only an absolute error bound, which today's tools report and which
may suffer from unnecessary over-approximations.


We realize that while this approach provides a practical solution, it is still
preliminary and can be improved in several ways.

First, a smarter subdivision strategy would be beneficial.
Currently, we divide the domain into equal-width intervals, and vary only their
number. The more fine-grained the subdivision, the bigger part of the domain can
be covered by relative error computations, however the running time increases
correspondingly. Ideally, we could exclude from the relative error computation
only a small domain around the zeros of the function $f$. While for univariate
functions, this approach is straight-forward (zeros can be, for instance,
obtained with a nonlinear decision procedure), for multivariate functions this
is challenging, as the zeros are not simple points but curves.

Secondly, subdivision could only be used for determining which sub-domains exhibit
potential division by zero. The actual relative error bound computation can then
be performed on the remainder of the input domain without subdividing it. This
would lead to performance improvements, even though the `guaranteed-no-zero'
domain can still consist of several disconnected parts.
Again, for univariate functions this is a straight-forward extension,
but non-trivial for multivariate ones.

Finally, we could compute
\begin{equation}
\max_{x_j\in I_{jk}}\left|\frac{f(x)-\tlf(x,e,d)}{f(x)+\epsilon}\right|
\label{eq:rel-via-epsilon}
\end{equation}
for some small $\epsilon$, which is a standard approach in scientific computing.
It is not sound, however, so that we do not consider it here.

\section{Experimental Evaluation}\label{sec:evaluation}
\renewcommand{\topfraction}{0.9}
    \setcounter{topnumber}{2}
    \setcounter{bottomnumber}{2}
    \setcounter{totalnumber}{4}     
    \setcounter{dbltopnumber}{2}    
    \renewcommand{\dbltopfraction}{0.9} 
    \renewcommand{\textfraction}{0.07}  

 \begin{table*}[t]
 \centering
 \renewcommand{\arraystretch}{1.1}
 \caption{Relative error bounds computed by different techniques}
 \label{tab:eval:nozero}
 \footnotesize{
 \begin{tabular}{ccccc|cc|cccc}
  \toprule
  & \multirow{2}{*}{\myhead{1.5cm}{Bench-\\mark}} & \multirow{2}{*}{\myhead{1.6cm}{Under-\\approx}} & \multirow{2}{*}{\textbf{Daisy}} & \multirow{2}{*}{\textbf{FPTaylor}} & \multirow{2}{*}{\myhead{1.5cm}{Naive\\approach}} & \multirow{2}{*}{\myhead{1.5cm}{Daisy\\+ subdiv}} &
  \multicolumn{4}{c}{\bf{DaisyOPT}}\\
   & & & & & & & \bf{Z3} & \bf{dReal} & \bf{Z3+subdiv} & \bf{dReal+subdiv} \\
  \midrule
  \parbox{0.09cm}{\multirow{7}{*}{\rotatebox[origin=c]{90}{Univariate}}} & bspline0 & 1.46e-15 & 4.12e-13 & 4.26e-13 & 5.11e+02 & 7.44e-14 & \textbf{3.00e-15} & \textbf{3.00e-15} & \textbf{3.00e-15} & \textbf{3.00e-15} \\
 & bspline1 & 7.91e-16 & \textbf{2.54e-15} & 3.32e-15 & 4.16e-01 & 5.32e-15 & 3.22e-15 & 3.22e-15 & 3.22e-15 & 3.22e-15 \\
  & bspline2 & 2.74e-16 & 1.11e-15 & 1.16e-15 & 5.22e-01 & 1.61e-15 & \textbf{8.92e-16} & 9.76e-16 & \textbf{8.92e-16} & \textbf{8.92e-16} \\
  & bspline3 & 5.49e-16 & 2.46e-10 & 3.07e-10 & 5.12e+05 & 5.23e-11 & \textbf{6.66e-16} & \textbf{6.66e-16} & \textbf{6.66e-16} & \textbf{6.66e-16} \\
  & sine & 2.84e-16 & 8.94e-16 & 8.27e-16 & 4.45e-01 & 1.39e-15 & \textbf{7.66e-16} & \textbf{7.66e-16} & \textbf{7.66e-16} & \textbf{7.66e-16} \\
 & sineOrder3 & 3.65e-16 & 1.04e-15 & 1.10e-15 & 1.39e-01 & 1.99e-15 & \textbf{8.94e-16} & \textbf{8.94e-16} & \textbf{8.94e-16} & \textbf{8.94e-16} \\
  & sqroot & 4.01e-16 & 1.04e-15 & 1.21e-15 & 1.02e+00 & 2.20e-15 & \textbf{1.02e-15} & \textbf{1.02e-15} & \textbf{1.02e-15} & \textbf{1.02e-15} \\
  \midrule
 \parbox{0.09cm}{\multirow{14}{*}{\rotatebox[origin=c]{90}{Multivariate}}} & doppler & 1.06e-15 & 2.08e-04 & 6.13e-07 & 2.09e+08 & 2.60e-05 & \textbf{1.93e-13} & \textbf{1.94e-13} & \textbf{1.93e-13} & \textbf{1.94e-13} \\
 & himmilbeau & 8.46e-16 & 6.55e-13 & 7.89e-13 & 6.69e+02 & 9.81e-15 & 6.54e-13 & 1.98e-15 & 7.05e-15 & \textbf{1.99e-15} \\
 & invPendulum & 3.74e-16 & 2.09e-11 & 2.48e-11 & 1.64e+00 & 1.22e-11 & \textbf{1.21e-15} & \textbf{1.35e-15} & \textbf{1.21e-15} & \textbf{1.52e-15} \\
 & jet & 1.45e-15 & 9.26e-15 & 7.53e-15 & 3.87e+00 & 1.40e-13 & \textbf{4.47e-15} & 5.12e-15 & 6.03e-15 & 6.51e-15 \\
 & kepler0 & 4.39e-16 & 1.31e-12 & 1.64e-12 & 2.16e+03 & 3.63e-12 & 3.97e-12 & 2.39e-15 & \textbf{1.63e-15} & 2.64e-15 \\
 & kepler1 & 7.22e-16 & 2.17e-11 & 2.59e-11 & 7.93e+04 & 8.70e-13 & 3.80e-11 & \textbf{1.29e-15} & 2.85e-13 & 1.71e-15 \\
 & kepler2 & 5.28e-16 & 4.01e-10 & 5.65e-15 & 4.09e+05 & 1.35e-11 & 4.56e-10 & 2.42e-15 & 8.58e-12 & \textbf{2.26e-15} \\
 & rigidBody1 & 4.49e-16 & 8.77e-11 & 1.14e-10 & 1.55e+00 & 2.50e-11 & \textbf{9.78e-16} & 1.27e-15 & \textbf{9.78e-16} & 1.46e-15 \\
 & rigidBody2 & 5.48e-16 & 3.91e-12 & 4.73e-12 & 5.14e+03 & 1.77e-12 & \textbf{2.21e-15} & 2.33e-15 & \textbf{2.21e-15} & 2.96e-15 \\
 & \parbox{1.4cm}{traincar\_state8} & 2.72e-15 & 2.16e-13 & 2.69e-13 & 2.91e+02 & 2.16e-13 & \textbf{7.67e-14} & 2.72e-13 & \textbf{7.67e-14} & 2.50e-13 \\
 & \parbox{1.4cm}{traincar\_state9} & 8.11e-16 & 3.44e-13 & 4.31e-13 & 3.47e+02 & 1.91e-13 & \textbf{3.45e-14} & 4.15e-13 & \textbf{3.45e-14} & 2.38e-13 \\
 & turbine1 & 5.79e-16 & 6.47e-13 & 1.48e-13 & 4.16e+02 & 6.81e-13 & \textbf{2.06e-15} & 3.07e-15 & \textbf{2.06e-15} & 3.90e-15 \\
 & turbine2 & 1.03e-15 & 5.26e-15 & 4.25e-15 & 4.81e+00 & 1.66e-13 & \textbf{4.12e-15} & 4.30e-15 & \textbf{4.12e-15} & 4.33e-15 \\
 & turbine3 & 7.41e-16 & 3.52e-13 & 7.43e-14 & 2.13e+02 & 3.91e-13 & \textbf{1.91e-14} & \textbf{1.92e-14} & \textbf{1.91e-14} & \textbf{1.93e-14} \\
  \bottomrule
 \end{tabular}
 }\vspace{-10pt}
\end{table*}
 We compare the different strategies for relative error computation on a set of standard benchmarks with
 FPTaylor and the forward dataflow analysis approach from Rosa (now implemented
 in Daisy) as representatives of state-of-the-art tools. We do not compare to
 Fluctuat directly as the underlying error estimation technique based on forward
 analysis with affine arithmetic is very similar to Daisy's. Indeed experiments performed
 previously~\cite{fptaylor,Darulova2017} show only small differences in computed error bounds.
 We rather choose to implement interval subdivision within Daisy.

 All experiments are performed in double floating-point precision (the
 precision FPTaylor supports), although all techniques in Daisy are parametric
 in the precision. The experiments were performed on a desktop computer running
 Debian GNU/Linux 8 64-bit with a 3.40GHz i5 CPU and 7.8GB RAM.
 The benchmarks bsplines, doppler, jetEngine, rigidBody, sine, sqrt and turbine are
 nonlinear functions from~\cite{rosa}; invertedPendulum and the
 traincar benchmarks are linear embedded examples from~\cite{Darulova2013}; and
 himmilbeau and kepler are nonlinear examples from the
 Real2Float project~\cite{Magron2015}.

 \begin{table*}[t]
 \centering
 \renewcommand{\arraystretch}{1.1}
 \caption{Analysis time of different techniques for relative errors on
 benchmarks without division by zero}
 \label{tab:eval:running-time}
 \footnotesize{
 \begin{tabular}{ccc|cc|cccc}
  \toprule
  \multirow{2}{*}{\textbf{Benchmark}} & \multirow{2}{*}{\textbf{Daisy}} & \multirow{2}{*}{\textbf{FPTaylor}} & \multirow{2}{*}{\myhead{1.7cm}{Naive\\approach}} & \multirow{2}{*}{\myhead{1.5cm}{Daisy\\+ subdiv}} &
  \multicolumn{4}{c}{\bf{DaisyOPT}}\\
  & & & & & \bf{Z3} & \bf{dReal} & \bf{Z3 + subdiv} & \bf{dReal + subdiv} \\
  \midrule
 bsplines & 6s & 13s & 13m 25s & 0.34s & 20s & 25s & 27s & 30s \\
sines & 5s & 8s & 13m 45s & 0.42s & 1m 4s & 1m 21s & 1m 8s & 1m 9s \\
sqroot & 3s & 6s & 6m 4s & 0.15s & 14s & 12s & 14s & 14s \\
  \midrule
doppler & 5s & 2m 11s & 2m 14s & 1s & 1m 59s & 2m 35s & 2m 58s & 7m 28s \\
himmilbeau & 9s & 4s & 5m 30s & 0.36s & 1m 50s & 1m 16s & 6m 15s & 8m 5s \\
invPendulum & 3s & 5s & 1m 31s & 0.15s & 7s & 37s & 25s & 3m 54s \\
jet & 20s & 17s & 19m 35s & 7s & 30m 40s & 32m 24s & 45m 31s & 2 h 20m 49s \\
kepler & 37s & 39s & 14m 41s & 1s & 3m 27s & 16m 29s & 12m 20s & 27m 56s \\
rigidBody & 11s & 8s & 10m 4s & 0.39s & 30s & 1m 18s & 1m 26s & 8m 37s \\
traincar & 10s & 42s & 8m 15s & 1s & 1m 1s & 10m 43s & 4m 7s & 18m 35s \\
turbine & 11s & 28s & 17m 25s & 2s & 5m 29s & 11m 28s & 12m 30s & 42m 36s \\
\midrule
\textbf{total} & 1m 60s & 5m 1s & 1h 52m 28s & 13s & 46m 42s & 1h 18m 45s & 1h 27m 22s & 4h 19m 53s \\
  \bottomrule
 \end{tabular}
 }\vspace{-10pt}
\end{table*}

\subsection{Comparing Techniques for Relative Error Bounds} \label{sec:eval_nozero}
 To evaluate the accuracy and performance of the different approaches
 for the case when no division by zero occurs, we modify the standard
 input domains of the benchmarks whenever necessary such that the function
 ranges do not contain zero and all tools can thus compute a non-trivial
 relative error bound.

 \autoref{tab:eval:nozero} shows the relative error bounds
 computed with the different techniques and tools, and~\autoref{tab:eval:running-time}
 the corresponding analysis times. Bold marks the best result,
 i.e. tightest computed error bound, for each benchmark. The column
 `Underapprox' gives an (unsound) relative error bound obtained with dynamic
 evaluation on 100000 inputs; these values provide an idea of the true
 relative errors.
 The `Naive approach' column confirms that simplifications of the relative error
 expression are indeed necessary (note the exponents of the computed bounds).

 The last four columns show the error bounds when relative errors are computed
 directly using the optimization based approach (denoted `DaisyOPT') from~\autoref{sec:tay_approach},
 with the two solvers and with and without subdivisions. For subdivisions,
 we use $m=2$ for univariate benchmarks, $m=8$ for multivariate and $p=50$ for both as
 in our experiments these parameters showed a good trade-off between performance and accuracy.
 For most of the benchmarks we find that direct evaluation of relative errors
 computes tightest error bounds with acceptable analysis times. Furthermore,
 for most benchmarks Z3, resp. its nonlinear decision procedure, is able to
 compute slightly tighter error bounds, but for three of our benchmarks dReal
 performs significantly better, while the running times are comparable.

 Somewhat surprisingly, we note that interval subdivision has limited effect on accuracy when combined with direct relative error computation, while also increasing the running time significantly.

 Comparing against state-of-the-art techniques (columns Daisy and FPTaylor), which
 compute relative errors via absolute errors, we notice that the results are sometimes
 several orders of magnitude less accurate than direct relative error computation
 (e.g. six orders of magnitude for the bspline3 and doppler benchmarks).

\begin{table*}[t]
  \centering 

  \caption{Relative error scalability with respect to the size of the input domain}
\label{tab:eval_scalable}   
  \footnotesize{
    \begin{tabular}{r|ccr|ccr}
      \toprule      
      \multirow{2}{*}{\textbf{Benchmark}} & \multicolumn{3}{c|}{\textbf{via absolute errors}} & \multicolumn{3}{c}{\textbf{directly}}\\
      & small & large & ratio & small & large & ratio \\
      \midrule
bspline0 & 6.44e-15 & 4.12e-13 & \emph{ 64 } & 9.99e-16 & 3.00e-15 & \emph{ 3 } \\
bspline1 & 1.57e-15 & 2.54e-15 & \emph{ 2 } & 2.07e-15 & 3.22e-15 & \emph{ 2 } \\
bspline2 & 6.71e-16 & 1.11e-15 & \emph{ 2 } & 6.75e-16 & 8.92e-16 & \emph{ 1.32 } \\
bspline3 & 3.27e-13 & 2.46e-10 & \emph{ 750.96 } & 6.66e-16 & 6.66e-16 & \emph{ 1 } \\
sine & 7.44e-16 & 8.94e-16 & \emph{ 1.20 } & 6.77e-16 & 7.66e-16 & \emph{ 1.13 } \\
sineOrder3 & 5.70e-16 & 1.04e-15 & \emph{ 2 } & 4.81e-16 & 8.94e-16 & \emph{ 2 } \\
sqroot & 6.49e-16 & 1.04e-15 & \emph{ 1.61 } & 5.65e-16 & 1.02e-15 & \emph{ 1.81 } \\
  \midrule
 doppler & 1.48e-11 & 2.08e-04 & \emph{ 1.40e+07 } & 1.26e-15 & 1.93e-13 & \emph{ 153.48 } \\
himmilbeau & 1.21e-15 & 6.55e-13 & \emph{ 541.15 } & 7.78e-16 & 6.54e-13 & \emph{ 841.07 } \\
invPendulum & 2.96e-13 & 2.09e-11 & \emph{ 70.74 } & 1.21e-15 & 1.21e-15 & \emph{ 1 } \\
jet & 9.05e-15 & 9.26e-15 & \emph{ 1.02 } & 4.60e-15 & 4.47e-15 & \emph{ 0.97 } \\
kepler0 & 1.40e-15 & 1.31e-12 & \emph{ 934.97 } & 1.17e-15 & 3.97e-12 & \emph{ 3.39e+03 } \\
kepler1 & 1.47e-15 & 2.17e-11 & \emph{ 1.47e+04 } & 3.06e-15 & 3.80e-11 & \emph{ 1.24e+04 } \\
kepler2 & 4.28e-15 & 4.01e-10 & \emph{ 9.37e+04 } & 7.42e-15 & 4.56e-10 & \emph{ 6.15e+04 } \\
rigidBody1 & 1.40e-12 & 8.77e-11 & \emph{ 62.66 } & 9.75e-16 & 9.78e-16 & \emph{ 1 } \\
rigidBody2 & 2.00e-15 & 3.91e-12 & \emph{ 1.95e+03 } & 1.16e-15 & 2.21e-15 & \emph{ 1.90 } \\
traincar\_state8 & 6.93e-15 & 2.16e-13 & \emph{ 31.18 } & 1.64e-15 & 7.67e-14 & \emph{ 46.85 } \\
traincar\_state9 & 4.61e-15 & 3.44e-13 & \emph{ 74.67 } & 1.73e-15 & 3.45e-14 & \emph{ 19.96 } \\
turbine1 & 4.46e-14 & 6.47e-13 & \emph{ 14.50 } & 1.75e-15 & 2.06e-15 & \emph{ 1.18 } \\
turbine2 & 6.94e-16 & 5.26e-15 & \emph{ 7.57 } & 6.91e-16 & 4.12e-15 & \emph{ 5.96 } \\
turbine3 & 1.10e-13 & 3.52e-13 & \emph{ 3.20 } & 6.50e-15 & 1.91e-14 & \emph{ 2.94 } \\
   \bottomrule
    \end{tabular}
  }
  \vspace{0.6em}
\end{table*}

 The column `Daisy+subdiv' shows relative errors computed via absolute errors,
 using the forward analysis with subdivision (with the same parameters as
 above). Here we observe that unlike for the directly computed relative errors,
 interval subdivision is mostly beneficial.

\subsection{Scalability of Relative Errors}
\label{sec:eval-scale}

We also evaluate the scalability of the direct computation of relative errors
with respect to the size of the input domain, and compare it with the
scalability of computation of relative errors via absolute. Since the
magnitude of roundoff errors (both absolute and relative) depends on the
magnitude of input values, larger input domains cause larger roundoff errors.
But also the over- approximation of the errors grows together with the size of
input domain. Ideally, we want this over-approximation to grow as slowly as
possible.

For the experiments in~\autoref{tab:eval:nozero}, we use as \emph{large} input
domains as possible without introducing result ranges which include zero. Now
for our scalability comparison we also compute relative errors on \emph{small}
input domains. For that we modify the standard input domains of the benchmarks
such that the width of input intervals is reduced, while the function ranges
still do not contain zero. All experiments are performed with the Z3 solver
with a timeout set to 1 second. The relative errors are computed without
interval subdivision, since we noticed that it has a limited effect on
accuracy while increasing running times.

\autoref{tab:eval_scalable} presents relative error bounds computed for smaller
and larger input domains. Columns `small' and `large' show relative errors
computed on smaller and larger input domains respectively. Column `ratio'
presents a relation between the values for the large and small domains. This
relation characterizes the scalability of the approach, the smaller, the
better.

Comparing the numbers from the `ratio' columns we notice that for direct
computation ratio is significantly smaller than for the computation via
absolute errors. This means that the over-approximation committed by the
direct computation is smaller than the over-approximation committed by the
relative error computation via absolute errors. The most prominent example is
the \emph{doppler} benchmark, where the directly computed error relative grew
for the larger domain by two orders of magnitude, while the relative error
computed via absolute grew by seven orders of magnitude. Based on these
results we conclude that relative errors computed directly scale better than
relative errors computed via absolute with respect to the size of the input
domain.


\subsection{Handling Division by Zero}

 To evaluate whether interval subdivision is helpful when dealing with the inherent
 division by zero challenge, we now consider the standard benchmark set, with
 standard input domains. \autoref{tab:eval:div-by-zero} summarizes our results.
 We first note that division by zero indeed occurs quite often, as the
 missing results in the Daisy and FPtaylor columns show.

 %
 The last three columns show our results when using interval subdivision.
 Note that to obtain results on as many benchmarks as possible we had to change
 the parameters for subdivision to $m=8$ and $p=50$ for
 univariate and $m=4,p=100$ for multivariate benchmarks.
 The result consists of three values: the first value is the maximum relative error
 computed over the sub-domains where relative error was possible to compute; in
 the brackets we report the maximum absolute error for the sub-domains where
 relative error computation is not possible, and the integer is the amount of
 these sub-domains where absolute errors were computed.
 We only report a result if the number of sub-domains with division by zero is
 less than 80\% of the total amount of subdomains, as larger numbers would probably be impractical to be used
 within, e.g. modular verification techniques.
 Whenever we report '-' in the table, this means that division by zero occurred on too many or all subdomains.

 We observe that while interval subdivision does not provide us with a
 result for all benchmarks, it nonetheless computes information
 for more benchmarks than state-of-the-art techniques.

\begin{table*}[t]
 \centering
 \renewcommand{\arraystretch}{1.1}
 \caption{Relative error bounds computed by different techniques on
 standard benchmarks (with potential division by zero)}
 \label{tab:eval:div-by-zero}
 \footnotesize{
 \begin{tabular}{cll|lll}
  \toprule

  \multirow{2}{*}{\textbf{Benchmark}} & \multirow{2}{*}{\textbf{Daisy}} & \multirow{2}{*}{\textbf{FPTaylor}} & \multirow{2}{*}{\myhead{1.8cm}{{Daisy\\ + subdiv}}} &
  \multicolumn{2}{c}{\bf{DaisyOPT}}\\
  & & & & \bf{Z3 + subdiv} & \bf{dReal + subdiv}\\
  \midrule
bspline0 & - & - & 1.58e-01 (1.08e-18, 1) & 3.00e-15 (1.08e-18, 1) & 3.00e-15 (1.08e-18, 1) \\
bspline1 & - & 3.32e-15 & 2.80e-13 & 3.22e-15 & 3.22e-15 \\
bspline2 & - & 3.50e-15 & 9.20e-16 & 8.92e-16 & 8.92e-16 \\
bspline3 & - & - & 1.31e-14 (9.67e-19, 1) & 6.66e-16 (9.67e-19, 1) & 6.66e-16 (9.67e-19, 1) \\
sine & - & - & 1.07e-15 (2.00e-16, 2) & 7.02e-16 (2.02e-16, 2) & 7.02e-16 (2.02e-16, 2) \\
sineOrder3 & - & - & 2.29e-15 (3.10e-16, 2) & 8.94e-16 (3.17e-16, 2) & 8.94e-16 (3.17e-16, 2) \\
sqroot & - & - & 7.09e-15 (2.83e-14, 3) & 1.92e-15 (3.11e-14, 3) & 1.92e-15 (3.11e-14, 3) \\
  \midrule
doppler & 1.48e-11 & 4.99e-12 & 8.95e-13 & 1.26e-15 & 1.35e-15 \\
himmilbeau & - & - & 3.75e-14 (1.00e-12, 12) & 2.57e-14 (1.00e-12, 12) & 2.84e-15 (1.00e-12, 12) \\
invPendulum & - & - & 4.94e-15 (2.60e-14, 32) & 2.82e-15 (2.60e-14, 32) & 3.08e-15 (2.60e-14, 32) \\
jet & - & - & - & - & - \\
kepler0 & 4.35e-15 & 4.57e-15 & 2.38e-13 (8.08e-14, 49) & 2.16e-15 (7.92e-14, 49) & 3.88e-15 (8.32e-14, 49) \\
kepler1 & 1.33e-14 & 1.17e-14 & - & - & - \\
kepler2 & - & 4.21e-14 & - & - & - \\
rigidBody1 & - & - & 2.29e-14 (2.16e-13, 46) & 1.07e-15 (2.16e-13, 46) & 1.78e-15 (2.16e-13, 46) \\
rigidBody2 & - & - & 2.65e-12 (3.51e-11, 50) & 1.67e-15 (3.65e-11, 50) & 3.80e-15 (3.65e-11, 50) \\
traincar\_state8 & - & - & - & - & - \\
traincar\_state9 & - & - & - & - & - \\
turbine1 & 6.12e-14 & 1.18e-14 & 6.03e-15 & 1.75e-15 & 5.21e-15 \\
turbine2 & - & - & 5.64e-14 (3.65e-14, 25) & 2.74e-15 (1.20e-13, 25) & 6.97e-14 (3.90e-14, 25) \\
turbine3 & 1.52e-13 & 2.21e-14 & 2.77e-14 & 6.50e-15 & 6.71e-15 \\
  \bottomrule
 \end{tabular}
 }
 \vspace{-10pt}
\end{table*}
\section{Related work}

The goal of this work is an automated and sound static analysis technique for
computing tight relative error bounds for floating-point arithmetic. Most
related are current static analysis tools for computing \emph{absolute} roundoff
error bounds~\cite{rosa,fluctuat,fptaylor,Magron2015}.

Another closely related tool is Gappa~\cite{Daumas2010}, which computes both
absolute and relative error bounds in Coq. It appears
relative errors can be computed both directly and via absolute errors. The
\emph{automated} error computation in Gappa uses intervals, thus, a
computation via absolute errors will be less accurate than Daisy performs. The
direct computation amounts to the naive approach, which we have shown to work
poorly.

The direct relative error computation was also used in the context of
verifying computations which mix floating-point arithmetic and bit-level
operations~\cite{Lee2016}. Roundoff errors are computed using an optimization
based approach similar to FPTaylor's. Their approach is targeted to specific
low-level operations including only polynomials, and the authors do not use
Taylor's theorem. However, tight error estimates are not the focus of the
paper, and the authors only report that they use whichever bound (absolute or
relative) is better. 
we are not aware of any systematic evaluation of different approaches for
sound relative error bounds.


More broadly related are abstract interpretation-based static analyses which are
sound wrt. floating-point arithmetic~\cite{Chen2008,Jeannet2009}, some of which
have been formalized in Coq~\cite{Jourdan2015} These domains, however, do not
quantify the difference between the real-valued and the finite-precision semantics
and can only show the absence of runtime errors such division-by-zero or overflow.


Floating-point arithmetic has also been formalized in an SMT-lib~\cite{smtlibFP}
theory and solvers exist which include floating-point
decision procedures~\cite{Brain2013,smtlibFP}. These are, however, not
suitable for roundoff error quantification, as a combination with the theory of
reals would be at the propositional level only and thus not lead to useful results.

Floating-point arithmetic has also been formalized in theorem provers such as
Coq~\cite{Boldo2011} and HOL Light~\cite{Jacobsen2015}, and
some automation support exists in the form of verification condition generation
and reasoning about ranges~\cite{Linderman2010,Ayad2010}. Entire numerical
programs have been proven correct and accurate within
these~\cite{Boldo2013,Ramananandro2016}. While very tight error bounds can be
proven for specific computations~\cite{Graillat2014}, these verification efforts
are to a large part manual and require substantial user expertise in
both floating-point arithmetic as well as theorem proving. Our work focuses on a
different trade-off between accuracy, automation and generality.

Another common theme is to run a higher-precision program alongside the original
one to obtain error bounds by testing~\cite{Benz2012,Chiang2014,Lam2013,Panchekha2015}.
Testing has also been used as a verification method for optimizing
mixed-precision computations~\cite{Rubio-Gonzalez2013,Lam2013b}.
These approaches based on testing, however, only consider a limited number of
program executions and thus cannot prove sound error bounds.

\section{Conclusion}

We have presented the first experimental investigation into the suitability of
different static analysis techniques for sound accurate relative error estimation.
Provided that the function range does not include zero, computing relative
errors \emph{directly} usually yields error bounds which are (orders of magnitude) more
accurate than if relative errors are computed via absolute errors (as is
current state-of-the-art). Surprising to us, while interval subdivision is beneficial
for absolute error estimation, when applied to direct relative error computation
it most often does not have a significant effect on accuracy.

We furthermore note that today's rigorous optimization tools could be improved
in terms of reliability as well as scalability. Finally, while interval subdivision
can help to alleviate the effect of the inherent division by
zero issue in relative error computation, it still remains an open challenge.


	\bibliographystyle{IEEEtran}
	\bibliography{biblio}


\begin{table*}[t]
	\centering
	
	\caption{Comparison of different configurations for subdivision (no division by zero)}
	\label{tab:eval-subdiv-config}
	\small{
		\begin{tabular}{crc|ccc|ccc}
			\toprule
			& \multirow{2}{*}{\textbf{Benchmark}} & \multirow{2}{*}{\textbf{\# of vars}} & \multicolumn{3}{c|}{\textbf{$p= 50$}} & \multicolumn{3}{c}{\textbf{$p= 100$}} \\
			& & \textbf{$m=2$} & \textbf{$m=5$} &\textbf{$m=8$} & \textbf{$m=2$} & \textbf{$m=5$} &\textbf{$m=8$}\\
			\midrule
     \parbox{0.09cm}{\multirow{7}{*}{\rotatebox[origin=c]{90}{Univariate}}} &  bspline0	&	1	&	3.00e-15	&	3.00e-15	&	3.00e-15	&	3.00e-15	&	3.00e-15	&	3.00e-15	\\
& bspline1	&	1	&	3.22e-15	&	3.22e-15	&	3.22e-15	&	3.22e-15	&	3.22e-15	&	3.22e-15	\\
& bspline2	&	1	&	8.92e-16	&	8.92e-16	&	8.92e-16	&	8.92e-16	&	8.92e-16	&	8.92e-16	\\
& bspline3	&	1	&	6.66e-16	&	6.66e-16	&	6.66e-16	&	6.66e-16	&	6.66e-16	&	6.66e-16	\\
& sine	&	1	&	7.66e-16	&	7.66e-16	&	7.66e-16	&	7.66e-16	&	7.66e-16	&	7.66e-16	\\
& sineOrder3	&	1	&	8.94e-16	&	8.94e-16	&	8.94e-16	&	8.94e-16	&	8.94e-16	&	8.94e-16	\\
& sqroot	&	1	&	1.02e-15	&	1.02e-15	&	1.02e-15	&	1.02e-15	&	1.02e-15	&	1.02e-15	\\
			\midrule
\parbox{0.09cm}{\multirow{14}{*}{\rotatebox[origin=c]{90}{Multivariate}}} & doppler	&	3	&	1.93e-13	&	1.93e-13	&	1.93e-13	&	1.93e-13	&	1.93e-13	&	1.93e-13	\\
& himmilbeau	&	2	&	7.83e-14	&	1.10e-14	&	7.05e-15	&	7.83e-14	&	1.10e-14	&	\textbf{5.80e-15}	\\
& invPendulum	&	4	&	1.21e-15	&	1.21e-15	&	1.21e-15	&	1.21e-15	&	1.21e-15	&	1.21e-15	\\
& jet	&	2	&	\textbf{4.47e-15}	&	4.92e-15	&	6.03e-15	&\textbf{4.47e-15}&	4.92e-15	&	\textbf{4.47e-15}	\\
& kepler0	&	6	&	3.66e-13	&	6.63e-13	&	\textbf{1.63e-15}	&	3.80e-13	&	7.05e-13	&	7.33e-13	\\
& kepler1	&	4	&	8.53e-12	&	8.10e-13	&	\textbf{2.85e-13}	&	8.33e-12	&	8.10e-13	&	4.63e-13	\\
& kepler2	&	6	&	9.65e-11	&	1.46e-11	&	8.58e-12	&	9.65e-11	&	1.46e-11	&	\textbf{5.61e-12}	\\
& rigidBody1	&	3	&	9.78e-16	&	9.78e-16	&	9.78e-16	&	9.78e-16	&	9.78e-16	&	9.78e-16	\\
& rigidBody2	&	3	&	2.21e-15	&	2.21e-15	&	2.21e-15	&	2.21e-15	&	2.21e-15	&	2.21e-15	\\
& traincar\_state8	&	14	&	7.67e-14	&	7.67e-14	&	7.67e-14	&	7.67e-14	&	7.67e-14	&	7.67e-14	\\
& traincar\_state9	&	14	&	3.45e-14	&	3.45e-14	&	3.45e-14	&	3.45e-14	&	3.45e-14	&	3.45e-14	\\
& turbine1	&	3	&	2.06e-15	&	2.06e-15	&	2.06e-15	&	2.06e-15	&	2.06e-15	&	2.06e-15	\\
& turbine2	&	3	&	4.12e-15	&	4.12e-15	&	4.12e-15	&	4.12e-15	&	4.12e-15	&	4.31e-15	\\
& turbine3	&	3	&	1.91e-14	&	1.91e-14	&	1.91e-14	&	1.91e-14	&	1.91e-14	&	1.91e-14	\\
			\bottomrule
		\end{tabular}
	}
	\vspace{0.6em}
\end{table*}


	\appendix[Experimental Comparison of Subdivision Parameters]
%

Our subdivision approach is parametric in the amount of subdivisions for each
interval $m$ and the total amount of optimizations $p$. In this appendix, we
present our experimental findings of good default configuration for interval
subdivision parameters ($m$ and $p$). We expect that the presence of zeros in
the function range might affect the choice of good default configuration.
Thus, we perform two separate experimental evaluations: on the input domains
where function range does not include zero~(\autoref{sec:eval-config}) and on
the standard input domains, where function range potentially contains
zero~(\autoref{sec:eval-config-divbyzero}).

\subsection{No Potential Division by Zero}
\label{sec:eval-config} 

To define good defaults for the subdivision parameters for the input domains
such that function range does not include zero, we have executed multiple
tests. During the experiments we compute relative errors directly for several
subdivision values $m=2,5,8$ in combination with two different values for the
limit for the total analysis runs number ($p = 50,100$). We used the Z3 solver
with a timeout of 1 second for testing all configurations.

\autoref{tab:eval-subdiv-config} shows the relative error bounds for different
subdivision values,~\autoref{tab:eval_configs_time} gives running times for
these configurations. The column '\# of vars' shows the amount of input variables
for each benchmark. Columns '2', '5', '8' show the bounds computed with $m=2,
5$ and $8$ subdivisions for input intervals respectively. 

\begin{table*}[t]
	\centering

		\caption[Comparison of running times for different configurations for subdivision (no division by zero)]{Comparison of running times for different configurations for subdivision}
		\label{tab:eval_configs_time}
		\footnotesize{
			\begin{tabular}{r|rrr|rrr}
				\toprule
				\multirow{2}{*}{\textbf{Benchmark}} & \multicolumn{3}{c|}{\textbf{$p= 50$}}& \multicolumn{3}{c}{\textbf{$p=100$}} \\
				& \textbf{$m=2$} & \textbf{$m=5$} &\textbf{$m=8$} & \textbf{$m=2$} & \textbf{$m=5$} &\textbf{$m=8$}\\
				\midrule
				bspline0		&	6s	&	12s	&	18s	&	7s	&	12s	&	17s	\\
				bspline1		&	8s	&	16s	&	24s	&	9s	&	16s	&	24s	\\
				bspline2		&	10s	&	18s	&	26s	&	11s	&	18s	&	27s	\\
				bspline3		&	2s	&	6s	&	8s	&	3s	&	5s	&	9s	\\
				sine		&	1m 3s	&	1m 13s	&	1m 26s	&	1m 19s	&	1m 14s	&	1m 41s	\\
				sineOrder3		&	5s	&	10s	&	15s	&	5s	&	10s	&	15s	\\
				sqroot		&	14s	&	24s	&	36s	&	15s	&	25s	&	35s	\\
				\midrule
				doppler		&	2m 41s	&	4m 6s	&	2m 58s	&	2m 44s	&	4m 7s & 7m 17s \\
				himmilbeau		&	2m 9s	&	8m 34s	&	6m 15s	&	2m 14s	& 8m 30s & 14m 59s \\
				invPendulum		&	52s	&	1m 7s	&	25s	&	55s	&	1m 6s & 2m 40s \\
				jet		&	351s	&	58m 45s	&	45m 31s	&	34m 1s	&	47m 57s & 1h 22m 35s \\
				kepler0		&	12m 47s	&	11m 25s	&	1m 40s	&	32m 39s	&	11m 49s & 21m 34s \\
				kepler1		&	15m 54s	&	16m 54s	&	5m 8s	&	15m 29s	&	17m 15s & 43m 8s \\
				kepler2		&	12m 7s	&	15m 46s	&	5m 32s	&	41m 41s	&	1h 16m 50s & 430s \\
				rigidBody1		&	30s	&	1m 17s	&	29s	&	30.74s	&	1m 16s & 2m 55s \\
				rigidBody2		&	1m 8s	&	2m 34s	&	57s	&	1m 7s	&	2m 35s & 5m 57s \\
				traincar\_state8		&	8m 10s	&	5m 57s	&	2m 13s	&	15m 39s	&	5m 57s & 14m 52s \\
				traincar\_state9		&	7m 28s	&	5m 12s	&	1m 54s	&	14m 20s	&	5m 26s & 13m 33s \\
				turbine1		&	2m 48s	&	4m 54s	&	2m 59s	&	2m 27s	&	4m 59s & 153s \\
				turbine2		&	4m 21s	&	16m 8s	&	6m 7s	&	3m 52s	&	15m 4s & 31m 57s \\
				turbine3		&	3m 17s	&	7m 9s	&	3m 24s	&	2m 57s	&	7m 14s & 14m 34s \\
				\midrule
				\textbf{total} & 1h 46m 49s & 2h 42m 26s & \textbf{1h 29m 7s} & 2h 52m 45s & 3h 32m 46s & 5h 11m 14s \\
				\bottomrule
			\end{tabular}
		}		
\vspace{-10pt}
\end{table*}

\begin{table*}[t]
	\centering

	\caption[Comparison of different configurations for subdivision (with potential division by zero)]{Comparison of different configurations for subdivision }
	\label{tab:divbyzero-subdiv-config}

	\footnotesize{
		\begin{tabular}{rc|ccc|ccc}
			\toprule
			\multirow{2}{*}{\textbf{Benchmark}} & \multirow{2}{*}{\textbf{\# of vars}} & \multicolumn{3}{c|}{\textbf{$p= 50$}} & \multicolumn{3}{c}{\textbf{$p= 100$}} \\
			& & \textbf{$m=4$} & \textbf{$m=6$} &\textbf{$m=8$} & \textbf{$m=4$} & \textbf{$m=6$} &\textbf{$m=8$}\\
			\midrule
bspline0 & 1 & 1 (4) & 1 (6) & 1 (8)& 1 (4) & 1 (6) & 1 (8)\\
bspline1 & 1 & 1 (4) & 1 (6) & 0 (8)& 1 (4) & 1 (6) & 0 (8)\\
bspline2 & 1 & 0 (4) & 0 (6) & 0 (8)& 0 (4) & 0 (6) & 0 (8)\\
bspline3 & 1 & 1 (4) & 1 (6) & 1 (8)& 1 (4)  & 1 (6) & 1 (8)\\
sine & 1 & 2 (4) & 2 (6) & 2 (8)& 2 (4) & 2 (6) & 2 (8)\\
sineOrder3 & 1 & - & 2 (6) & 2 (8)& - & 2 (6) & 2 (8)\\
sqroot & 1 & 3 (4) & 2 (6) & 3 (8)& 3 (4) & 2 (6) & 3 (8)\\
			\midrule
   doppler & 3 & 0 (16)& 0 (36) & 0 (8)& 0 (64) & 0 (36) & 0 (64) \\
himmilbeau & 2 & 12 (16)& 17 (36) & - & 12 (16)& 17 (36) & 15 (64) \\
invPendulum & 4 & 12 (16)& 22 (36) & 6 (8)& 32 (64) & 22 (36) & 34 (64) \\
jet & 2 & \underline{15 (16)} & \underline{33 (36)} & - & \underline{15 (16)} & \underline{33 (36)} & \underline{56 (64)} \\
kepler0 & 6 & - & - & - & 49 (64) & - & - \\
kepler1 & 4 & - & - & - & \underline{63 (64)} & - & - \\
kepler2 & 6 & - & - & - & - & - & - \\
rigidBody1 & 3 & - & - & - & 46 (64) & - & - \\
rigidBody2 & 3 & - & - & - & 50 (64) & - & - \\
traincar\_state8 & 14 & - & - & - & - & - & - \\
traincar\_state9 & 14 & - & - & - & - & - & - \\
turbine1 & 3 & 0 (16)& 0 (36) & 0 (8)& 0 (64) & 0 (36) & 0 (64) \\
turbine2 & 3 & 10 (16)& 17 (36) & 6 (8)& 25 (64) & 17 (36) & 25 (64) \\
turbine3 & 3 & 0 (16)& 0 (36) & 0 (8)& 0 (64) & 0 (36) & 0 (64) \\
			\bottomrule
		\end{tabular}
	}
\end{table*}
We noticed that, perhaps surprisingly, a more fine-grained subdivision does
not gain accuracy for most of the benchmarks. For those benchmarks where we
observed different error bounds three out of five tightest bounds
(\emph{himmilbeau}, \emph{jet} and \emph{kepler2}) were computed with the
configuration $m=8, p=100$. The second best choice in terms of accuracy is
$m=8, p=50$, when the bounds for \emph{himmilbeau}, \emph{jet} and
\emph{kepler} differ from the tightest computed bounds only insignificantly.

One may think that the greater the number of subdivisions $m$ is, the larger
are the running times. Indeed it is the case when \emph{all} input intervals
are subdivided. Recall that, to limit this effect we introduced the total
amount of optimizations $p$. For the univariate benchmarks the influence of
$p$ is not noticeable, since the total amount of optimizations is equal to the
amount of subdivisions in this case, and is below the limit $p$. For
multivariate benchmarks we observe that $p$ limits the amount of variables
where intervals are subdivided and hence limits runtime. Thus, the
configuration with the \emph{greatest} tested \emph{subdivision} value $8$ has
the \emph{smallest} total \emph{running time} when $p=50$.

Keeping in mind the accuracy-performance trade-off, we select the following
default configuration: for univariate benchmarks $m=2$ and $p=50$; for
multivariate benchmarks $m=8$ and $p=50$.

\subsection{With Potential Division by Zero}
\label{sec:eval-config-divbyzero}

To find good default subdivision configuration for relative error computation
with potential division by zero, we recall semantic of each parameter. The
parameter $m$ regulates how fine-grained the subdivision of each input
interval should be, while the parameter $p$ is intended to limit the running
time. $p$ bounds not only the total amount of optimization runs, but also
regulates for how many variables the input intervals are subdivided. This
balance between more \emph{subdivisions for one} interval and more input
\emph{intervals being subdivided} may change if we have to deal with potential
division by zero. Thus, we do not reuse the configuration found
in~\autoref{sec:eval-config}, and perform the comparison for several values of
$m$ and $p$. We compare results for $m=4,6$ and $8$ and $p=50, 100$.

The focus of this comparison is to find a configuration which computes
relative errors for as many benchmarks as possible and for as big part of the
input domain as possible. Therefore, for different configurations we compare
the amount of sub-domains where computations failed because of division by
zero. ~\autoref{tab:divbyzero-subdiv-config} summarizes our results. Columns
$m=4$, $m=6$ and $m=8$ show the amount of sub-domains where division by zero
occurred for $4$, $6$ and $8$ subdivisions for each input interval with upper
limit for total amount of optimizations $p=50$ and $100$ respectively.  The
result consists of two values: the first value is the amount of sub-domains
where computation of relative errors failed, the second is the total amount of
sub-domains, e.g. $0 (4)$ means that relative error has been successfully
computed on all sub-domains. Underline marks the results, for which relative
error computation failed (due to division by zero) on more than $80\%$ of sub-
domains. We consider such results to be impractical and thus will not report
error bounds for these cases. Whenever we report `-' in the table, this means
that relative error computations reported division by zero for all sub-
domains.

For univariate benchmarks we see for almost all benchmarks all configurations
provided relative error bounds. Only for \emph{sineOrder} subdivision of $m=4$
sub-intervals is not sufficient to obtain a result. Interestingly, for
\emph{bspline1} computations reported division by zero for $m=4$ and $m=6$,
but for the more fine-grained subdivision $m=8$ no division by zero occurred,
and it was possible to compute relative error bound for all sub-domains. Since
the configuration $m=8$ $p=50$ allows to compute relative error for all
univariate benchmarks in our set, we choose it as a default for univariate
benchmarks. Note that for univariate benchmarks it does not play a role
whether we take $p=50$ or $p=100$, as the amount of sub-domains for all tested
values of $m$ is lower than $50$ and $100$.

We noticed that for multivariate benchmarks there is one configuration that
allowed to compute relative error for most of the benchmarks, that is $m=4$,
$p=100$. That means that for this set of multivariate benchmarks it is
beneficial to subdivide each individual interval less, while having intervals
for more variables subdivided. We choose this configuration as default for
multivariate benchmarks.

For some benchmarks, however, independent from the subdivision parameters it
was still not possible to compute any estimate of relative error, or the
computed error bound is valid for only a small sub-domain (\emph{jet} and
\emph{kepler1}).

\end{document}